\documentclass[twocolumn,preprintnumbers,amsmath,amssymb,superscriptaddress]{revtex4}
\usepackage[dvipdfmx]{graphicx}
\usepackage{color}
\usepackage{dcolumn}
\usepackage{bm}

\begin{document}

\title{Oxidation annealing effects on the spin-glass-like magnetism and appearance of superconductivity in T*-type La$_{1-x/2}$Eu$_{1-x/2}$Sr$_x$CuO$_4$ (0.14 $\leq x \leq$ 0.28)}



\author{Shun Asano}
\email{shun.asano@imr.tohoku.ac.jp}
\affiliation{Department of Physics, Tohoku University, Aoba, Sendai 980-8578, Japan}
\affiliation{Institute for Materials Research, Tohoku University, Katahira, Sendai 980-8577, Japan}
\author{Kensuke M. Suzuki}
\affiliation{Institute for Materials Research, Tohoku University, Katahira, Sendai 980-8577, Japan}
\author{Kota Kudo}
\affiliation{Department of Physics, Tohoku University, Aoba, Sendai 980-8578, Japan}
\author{Isao Watanabe}
\affiliation{Advanced Meson Science Laboratory, Nishina Center for Accelerator-Based Science, The Institute of Physical and Chemical Research (RIKEN), Wako, Saitama 351-0198, Japan}
\author{Akihiro Koda}
\author{Ryosuke Kadono}
\affiliation{Institute of Materials Structure Science, High Energy Accelerator Research Organization, Tsukuba, Ibaraki 305-0801, Japan}
\author{Takashi Noji}
\author{Yoji Koike}
\affiliation{Department of Applied Physics, Graduate School of Engineering, Tohoku University, Sendai 980-8579, Japan}
\author{Takanori Taniguchi}
\author{Shunsaku Kitagawa}
\author{Kenji Ishida}
\affiliation{Department of Physics, Kyoto University, Kyoto 606-8502, Japan}
\author{Masaki Fujita}
\email{fujita@imr.tohoku.ac.jp}
\affiliation{Institute for Materials Research, Tohoku University, Katahira, Sendai 980-8577, Japan}
\affiliation{Institute of Materials Structure Science, High Energy Accelerator Research Organization, Tsukuba, Ibaraki 305-0801, Japan}

\begin{abstract}
We investigated the magnetism and superconductivity in as-sintered (AS) and oxidation annealed (OA) T*-type La$_{1-x/2}$Eu$_{1-x/2}$Sr$_x$CuO$_4$ (LESCO) with 0.14 $\leq x \leq$ 0.28 by the first comprehensive muon spin rotation/relaxation ($\mu$SR), magnetic susceptibility, and electrical resistivity measurements. In OA superconducting samples, no evidence of magnetic order was observed, whereas AS semiconducting samples exhibited evidence of a disordered magnetic state in the measured temperature range between $\sim$4 K and $\sim$8 K. 
Therefore, the ground state in LESCO drastically varies with oxidation annealing and the magnetic phase competitively exists with the superconducting (SC) phase. 
The magnetic phase in the AS LESCO is quite robust against Sr doping, while the SC phase degrades with increasing $x$. A monotonous decrease of the SC transition temperature from 24.5 K in $x$ = 0.14 to 9.0 K in $x$ = 0.28 suggests the disappearance of the SC phase at $x$ $\sim$ 0.34. 
Furthermore, we clarified the simultaneous development of (quasi) static magnetism and the electrical resistivity at a low temperature in AS samples, suggesting the inducement of magnetism by the suppression of carrier mobility. The variation in magnetism due to annealing is discussed from a viewpoint of structural defects, which was previously reported from neutron diffraction measurements. 

\end{abstract}

\maketitle

\section{Introduction}

In the research of high superconducting-transition-temperature ($T_{\rm c}$) superconductivity in copper oxides, the effect of the crystal structure on their physical properties is one of the fundamental issues that need to be studied.
Recently, the ground state of the parent $R_{2}$CuO$_4$ ($R$CO) has been discussed from a viewpoint of Cu coordination, that is, the existence/absence of apical oxygens above and below Cu ions. Naito's group first confirmed a superconducting (SC) transition in a thin film of $R$CO with a Nd$_{2}$CuO$_4$-type (T'-type) structure having CuO$_4$ coplanar coordination~\cite{Tsukada2005}, which has been regarded as a Mott insulator for a long time. 
Superconductivity was subsequently observed by Takamatsu {\it et al} in low-temperature-synthesized polycrystalline samples of parent T'-type La$_{1.8}$Eu$_{0.2}$CuO$_4$\cite{Takamatsu2012}. 
From the theoretical aspect, it was shown by an ab-initio calculation conducted by Weber {\it et al} that the T'-type compound can have the metallic ground state, while the ground state of K$_2$NiF$_4$-type (T-type) cuprates containing CuO$_6$ octahedra is Mott insulating state.\cite{Weber2010} 
Furthermore, the decrease of electron correlation strengths with increasing the distance between apical oxygen and the CuO$_2$ plane was demonstrated~\cite{Jang2015a}, supporting the Slater picture for T'-type. 
Partially existing apical oxygens in the as-sintered (AS) T'-type compound are considered to prevent the appearance of superconductivity because of the introduction of random electronic potential in the CuO$_2$ plane~\cite{Adachi2017}. Thus, a reduction annealing procedure is necessary for the complete removal of excess oxygen to induce superconductivity in T'-type $R_2$CuO$_4$. 

For a further study on the relationship between Cu coordination and the physical properties, other reference systems are important. A single-layer T*-type cuprate with CuO$_5$ pyramid coordination, which is formed by alternate stacks of rock-salt layers in the T-type cuprate, and by fluorite layers in the T'-type one along the c-direction, is an isomer of $R_{2}$CuO$_4$. Hole-doped Nd$_{2-x-y}$Ce$_x$Sr$_y$CuO$_4$ (NCSCO) was reported as the first T*-type superconductor ($T_{\rm c} \sim 32$ K). \cite{Akimitsu1988, Sawa1989} The AS compound shows semiconducting behavior even in the heavily hole-doped region \cite{Izumi1989}, and oxidation annealing under high pressure is required for the emergence of superconductivity. The main role of annealing was reported to be the repair of oxygen vacancies~\cite{Izumi1989, Border1990}, which is an opposite method proposed to remove apical oxygens by reduction annealing in T'-type cuprates. 
Many T*-type cuprates with different compositions were synthesized after the discovery of SC NCSCO~\cite{Cheong89, Fisk89}. However, the physical properties of T*-type cuprates have been left unquestioned for a long time. This is because of the difficulties in controlling the crystal growth and in obtaining SC samples by high pressure heat treatment. Therefore, T*-type cuprates have been out of mainstream high-$T_{\rm c}$ research.

Although studies on electronic states in T*-type cuprates are quite limited, an angle-resolved photoemission spectroscopy (ARPES) measurement can clarify the electronic structure of T*-type superconductors\cite{Ino2004}. Different features in the ARPES spectra from those for underdoped La$_{2-x}$Sr$_x$CuO$_4$ (LSCO)~\cite{Lanzara2001} with larger magnitude of pseudogap in SmLa$_{0.85}$Sr$_{0.15}$CuO$_{4}$ suggests a distinct doping evolution of the electric structure between T*- and T-type cuprates. Thus, the investigation of doping effects on the magnetic properties is important. Regarding the magnetism of T*-type cuprates, the three-dimensional magnetic order in the undoped parent La$_{1.2}$Tb$_{0.8}$CuO$_4$ was observed by muon spin rotation/relaxation ($\mu$SR) measurements in the early stage of research\cite{Lappas94}. However, to the best of the authors' knowledge, until our recent $\mu$SR measurement on AS La$_{0.9}$Eu$_{0.9}$Sr$_{0.2}$CuO$_4$, no experimental study on magnetism of carrier-doped T*-type cuprates has been reported. Therefore, the magnetism in the SC phase and the doping evolution of spin correlations are still unknown.

In this paper, we revisited T*-type cuprates and elucidated the magnetic and SC properties comprehensively for the first time with the aim of complementing the lack of basic knowledge for the above relevant issues. Our preliminary $\mu$SR measurement on AS La$_{1-x/2}$Eu$_{1-x/2}$Sr$_x$CuO$_4$ (LESCO) with $x$ = 0.20 exhibited the development of magnetism at a low temperature\cite{Fujita2018}. Therefore, by extending the $\mu$SR study on LESCO, the effect of hole-doping and oxidation annealing on the inherent magnetism in T*-type cuprate can be clarified. 


\section{Sample preparation and experimental details}

The polycrystalline samples of LESCO with 0.14 $\leq x \leq$ 0.28 were synthesized by a solid-state reaction method. Dried powders of La$_2$O$_3$, Eu$_2$O$_3$, SrCO$_3$, and CuO were mixed with the nominal composition. The mixture was pressed into pellets and sintered at 1050 $^{\circ}$C in air with intermediate grindings. The oxygenized samples were prepared by annealing the AS samples in oxygen gas under 40 MPa at 500 $^{\circ}$C for 80 h. The phase purity of the samples was checked by X-ray powder diffraction measurements. Samples with $x >$ 0.14 were confirmed to be single phase T*-type cuprates, while  $x=0.14$ sample contained impurity phase with a volume fraction of $\sim$7 $\%$ of T'-type cuprate. The value of oxygen gain through annealing ($y$) was evaluated to be 0.023--0.024 per formula unit for $x$ = 0.18, 0.20, 0.22, and 0.26 from the variation of sample weight. Although the $y$ for other samples was not determined, we confirmed the increase of the c-lattice constant (decrease of a-lattice constant) in all samples after annealing, suggesting an introduction of oxygens.

	\begin{figure}[t]
	\begin{center}
	\includegraphics[width=83mm]{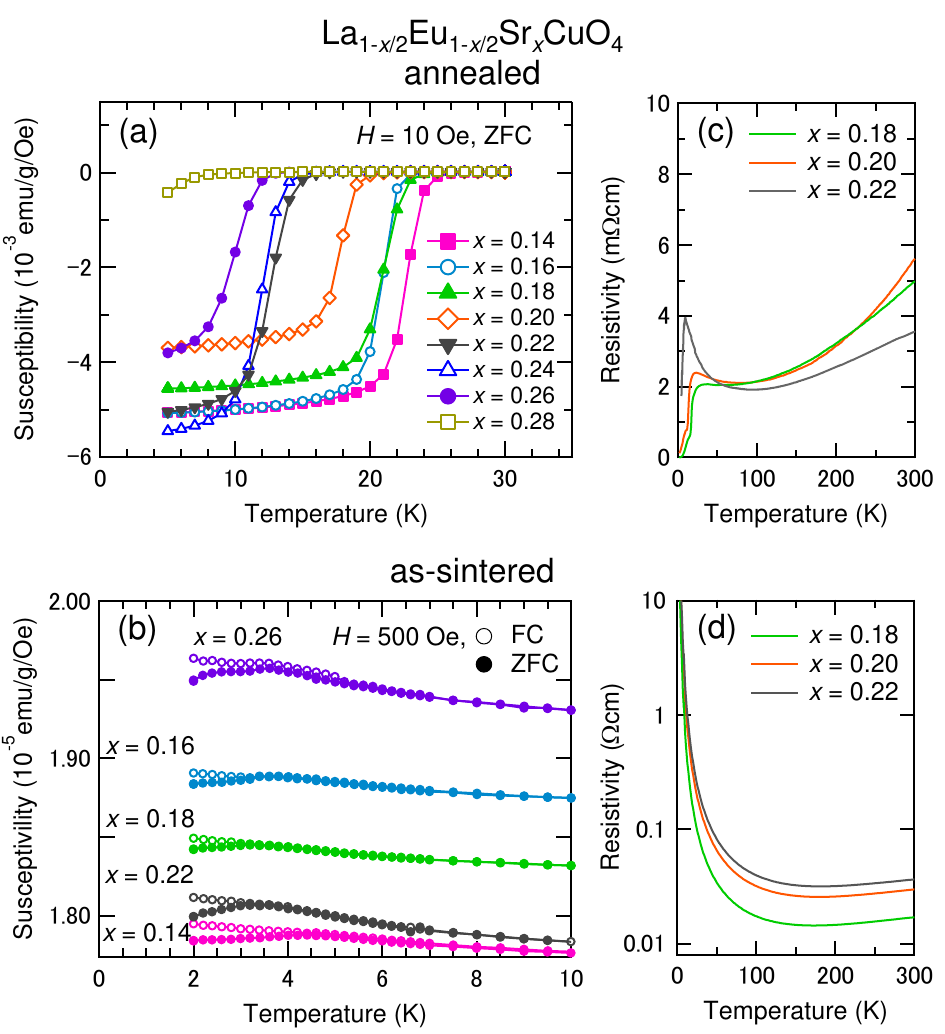}
	\caption{(Color online) (a) Magnetic susceptibility of oxidation annealed (OA) La$_{1-x/2}$Eu$_{1-x/2}$Sr$_x$CuO$_4$ with $x=0.14$--0.28 in a magnetic field of 10 Oe after zero field cooling, and (b) that of as-sintered (AS) ones measured under zero-field-cooled (closed  circles) and field-cooled (open circles) processes with a field of 500 Oe. Electrical resistivity for (c) OA and (d) AS La$_{1-x/2}$Eu$_{1-x/2}$Sr$_x$CuO$_4$ with $x=0.18$, 0.20, and 0.22.}
	\label{Chi_Resistivity_v4}
	\end{center}
	\end{figure}

Magnetic susceptibility was measured for AS and OA LESCO using a superconducting quantum interference device. We furthermore performed electrical resistivity measurements for $x$ = 0.18, 0.20, and 0.22 samples with a standard dc four-probe method. Zero-field (ZF) and longitudinal-field (LF) ${\mu}$SR measurements were carried out using a pulsed positive muon beam on the D1 and S1 (ARTEMIS) spectrometers in the Materials and Life Science Experimental Facility (MLF) in J-PARC, Japan, and the spectrometer (CHRONUS) at Port 4 in the RIKEN-RAL Muon Facility (RAL) in the Rutherford Appleton Laboratory, UK.
The $\mu$SR measurements for AS LESCO with $x = $0.20 and 0.28, and AN LESCO with $x=0.24$ were performed in MLF, and those for other samples were performed in RAL.
In most of the measurements, the samples were cooled down to 4 K using an open-cycle $^4$He-flow cryostat. Only the AS LESCO with $x=0.14$ was cooled down to 1.9 K using a cryostat different from that used for measurements above 4 K. The time ($t$) evolution of muon spin polarization ($\mu$SR time spectra, ($A$($t$))) after the implantation of muons into the sample was measured by the asymmetry of the decay positron emission rate between forward and backward counters.
All $\mu$SR time spectra in this paper are shown after correcting the counting efficiency and the solid angle of the detectors.

\section{Results}
\subsection{Magnetic susceptibility and electrical resistivity measurements}

Figures \ref{Chi_Resistivity_v4}(a) and 1(b) show the temperature dependences of magnetic susceptibility measured in a magnetic fields of 10 Oe and 500 Oe for OA and AS  LESCO with various $x$ from 0.14 to 0.28, respectively. The Meissner effect attributed to the appearance of the SC state was observed in all the OA samples, although the signal was weak in $x$ = 0.28. In contrast, in the AS samples, no evidence of SC transition was confirmed. However, as seen in Fig. \ref{Chi_Resistivity_v4}(b), the susceptibility measured in a magnetic field of 500 Oe under zero-field-cooled (ZFC) and field-cooled (FC) processes shows different behaviors at low temperature, which is typical for a spin-glass (SG) system. We evaluated the onset temperature for the appearance of the SC state ($T_{\rm c}$) in OA samples and the characteristic temperature at which the ZFC susceptibility in AS samples shows a local maximum ($T_{\rm sg}$). $T_{\rm c}$ and $T_{\rm sg}$ are summarized in Fig. \ref{phase_diagram_AN} and Fig. \ref{phase_diagram_AS_v2}(a) as a function of $x$.

	\begin{figure}[t]
	\begin{center}
	\includegraphics[width=78mm]{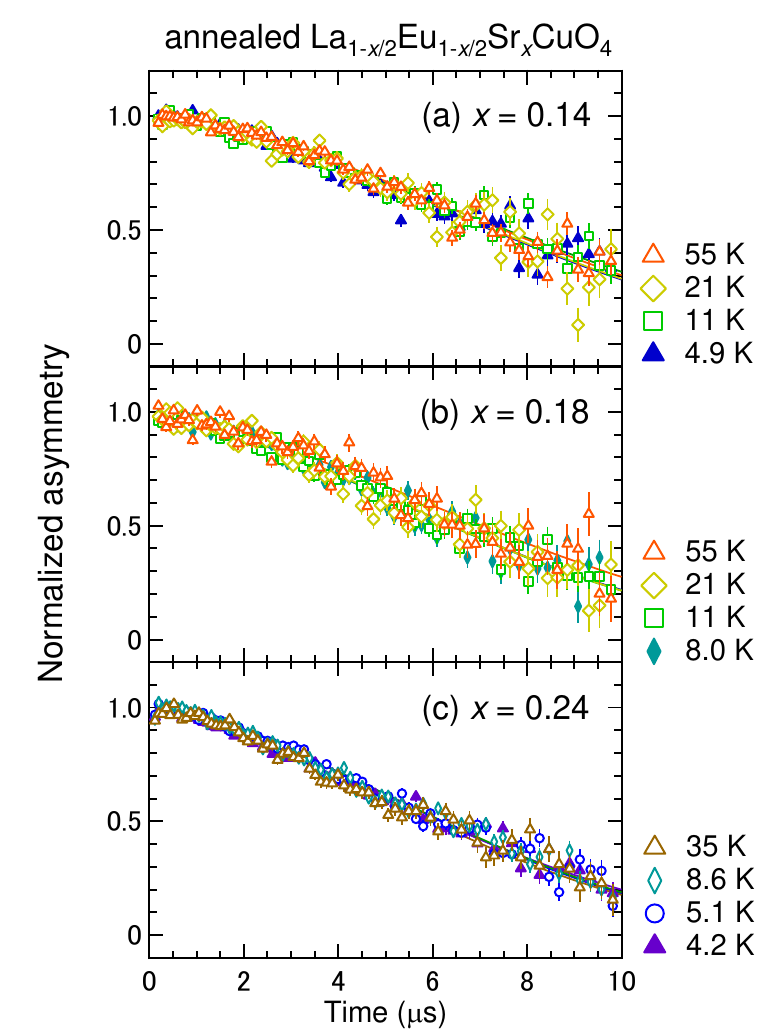}
	\caption{(Color online) Zero-field $\mu$SR time spectra of oxidation annealed La$_{1-x/2}$Eu$_{1-x/2}$Sr$_x$CuO$_4$ with (a) $x=0.14$ (RAL), (b) 0.18 (RAL), and (c) 0.24 (MLF). The solid curves are fitted results using the first term of Eq. (\ref{eq:Eq_1}).}
	\label{spectra_AN_v4}
	\end{center}
	\end{figure}
	
The temperature dependence of electrical resistivity ($\rho$) for OA and AS LESCO with $x$ = 0.18, 0.20, and 0.22 is shown in Figs. \ref{Chi_Resistivity_v4}(c) and \ref{Chi_Resistivity_v4}(d). The $\rho$ for all AS samples shows semiconducting behavior. Upon cooling, the $\rho$ slightly decreases and rapidly increases at low temperature, indicating the existence of carriers and the strong localization of carriers. In the OA samples, the $\rho$ exhibits metallic behavior above $\sim$100 K, while it shows an upturn and increases upon cooling, followed by a rapid drop associated with the emergence of superconductivity. 

	\begin{figure}[t]
	\begin{center}
	\includegraphics[width=78mm]{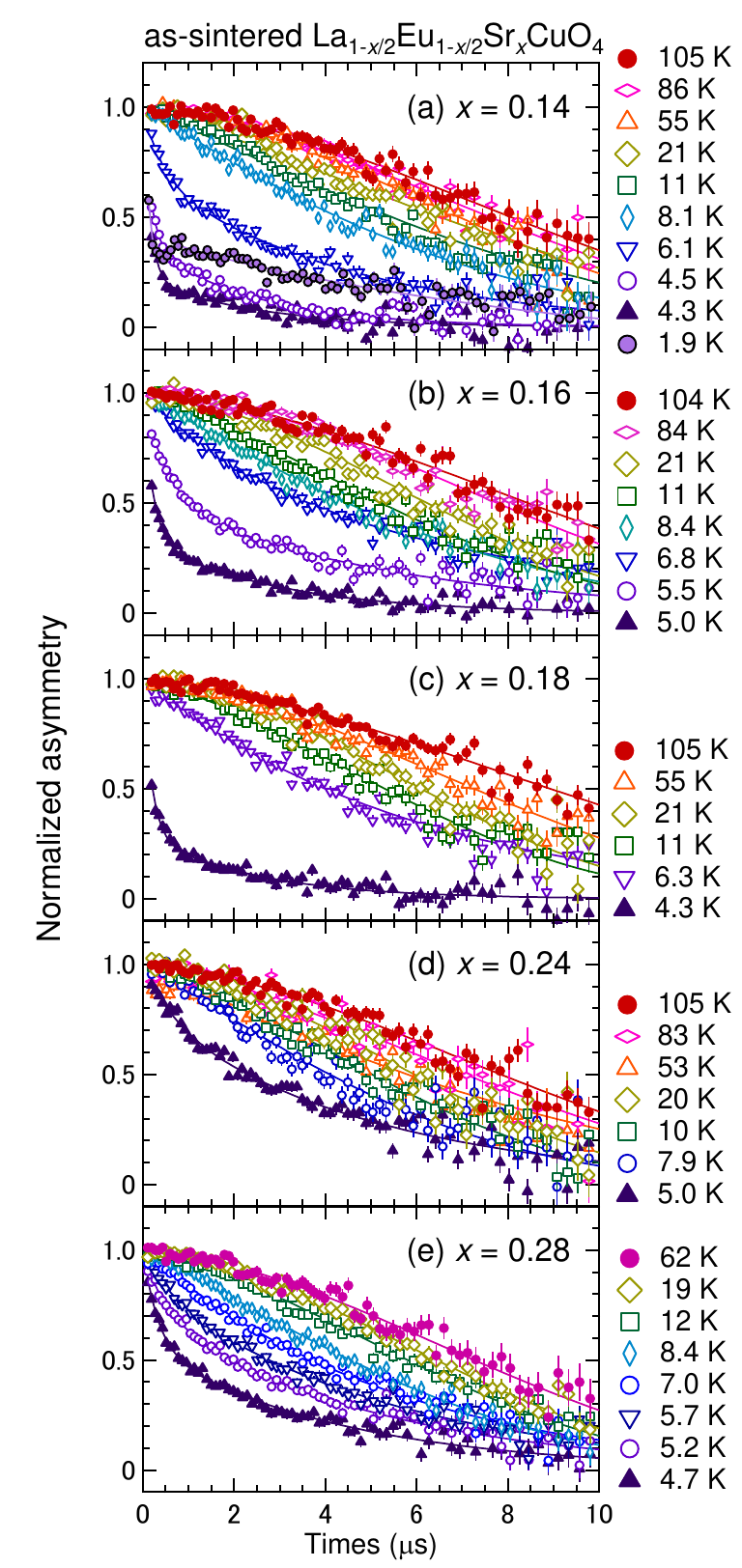}
	\caption{(Color online) Zero-field $\mu$SR time spectra of as-sintered La$_{1-x/2}$Eu$_{1-x/2}$Sr$_x$CuO$_4$ with (a) $x=0.14$ (RAL), (b) 0.16 (RAL), (c) 0.18(RAL), (d) 0.24 (RAL), and (e) 0.28 (MLF). The solid curves are fitted results using Eq. (\ref{eq:Eq_1}) for the data collected above 4 K.}
	\label{spectra_AS_v5}
	\end{center}
	\end{figure}

\subsection{Zero-field $\mu$SR measurements}
We first investigated the Cu spin dynamics in OA LESCO by ZF-$\mu$SR measurements. Figure \ref{spectra_AN_v4} shows normalized $\mu$SR time spectra for OA LESCO with $x=0.14$, 0.18, and 0.24. All spectra show Gaussian-type depolarization above 4 K and the thermal evolution of the spectra is negligible. This Gaussian-type depolarization is originated from the nuclear-dipole field, indicating neither static nor dynamic fluctuating Cu spins within the $\mu$SR time window. 
%
	\begin{figure}[tb]
	\begin{center}
	\includegraphics[width=82mm]{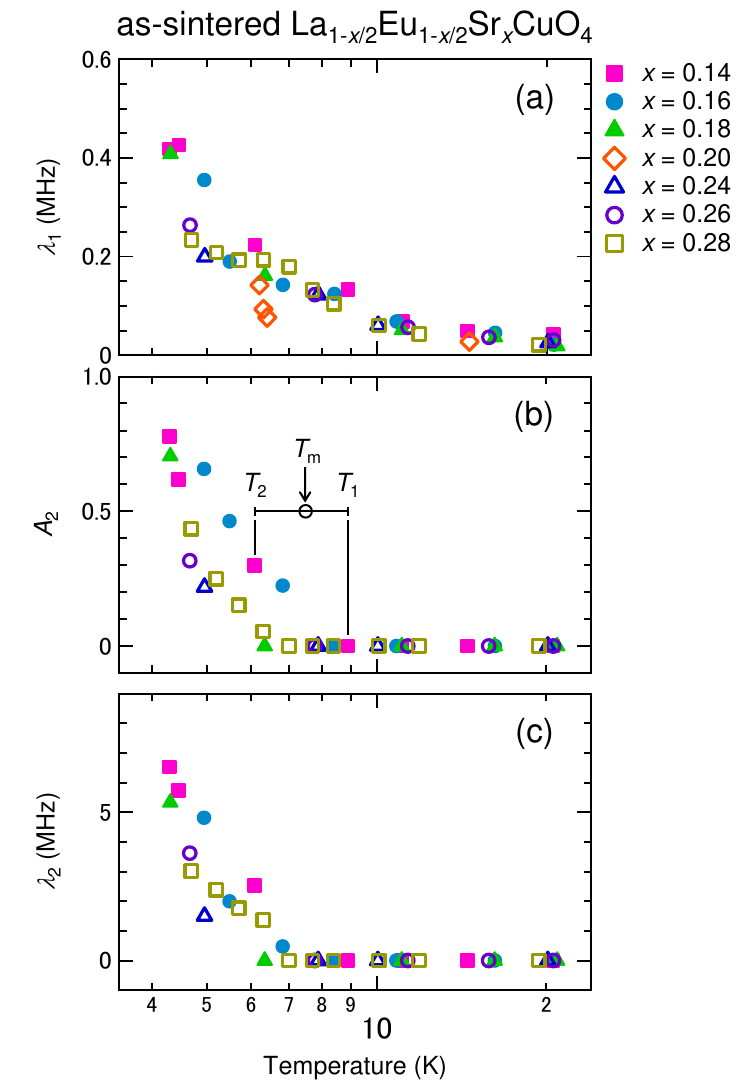}
	\caption{(Color online) Temperature dependence of (a) muon-spin depolarization rate $\lambda_{1}$ of the slow depolarizing component, (b) initial asymmetry $A_{\rm 2}$ and (c) depolarization rate $\lambda_{2}$ of the fast depolarizing component in Eq. (\ref{eq:Eq_1}) for as-sintered La$_{1-x/2}$Eu$_{1-x/2}$Sr$_x$CuO$_4$ ($0.14~{\leq}~x~{\leq}~0.28$).  $T_{\rm m}$ is the intermediate point between the lowest temperature where $A_{\rm 2}$ = 0 ($T_{\rm 1}$) and the first temperature where $A_{\rm 2}$ has a finite value upon cooling ($T_{\rm 2}$).}
	\label{spectra_AS_Prm_v6}
	\end{center}
	\end{figure}
In contrast to the absence of static magnetism in the OA samples, an appearance of (quasi) static magnetism was confirmed in AS non-SC LESCO samples with 0.14 $\leq x \leq$ 0.28. Figure \ref{spectra_AS_v5}(a) shows ZF-$\mu$SR time spectra in AS LESCO with $x=0.14$. Upon cooling from 105 K, the Gaussian-type depolarization changes into an exponential-type one, indicating the development of Cu spin correlations. At 1.9 K, the  spectrum shows oscillating behavior, indicating the existence of coherent magnetic order. 
The spectral change from Gaussian-type depolarization to exponential-type one was observed in all AS LESCO with 0.16 $\leq x \leq$ 0.28, including previously reported $x$ = 0.20\cite{Fujita2018}. 
Comparing the spectra at low temperature of 4--5 K for all samples shown in Fig. \ref{spectra_AS_v5}, the depolarization seems slightly slower with increasing $x$, suggesting a weak degradation of magnetism upon doping.

For the quantitative analysis of the $\mu$SR time spectra in AS LESCO above 4 K, we fitted the following function to the obtained data with taking a microscopic phase separation between magnetic and non-magnetic phases into account.
	\begin{equation}
	A(t)=A_{1}e^{-{{\it\lambda}_{1}}t}G({\it\Delta}, t)+A_{2}e^{-{{\it\lambda}_2}t} \nonumber, \hspace{12mm} (1)
	\label{eq:Eq_1}
	\end{equation}	
where $A_{1}+A_{2}=1$. 
The first (second) term represents the slow (fast) depolarizing component due to fast (slow) Cu fluctuations.
$G({\it \Delta}, t)$ is the static Kubo--Toyabe function and $\it \Delta$ is the half-width of the nuclear dipole field distributed at the muon site. $A_{\rm 1}$ ($A_{\rm 2}$) and $\lambda_1$ ($\lambda_2$) are initial asymmetry and depolarization rates of the slow (fast) depolarizing component, respectively. The spectra are well reproduced by Eq. (\ref{eq:Eq_1}) for all samples, as shown by solid curves in Fig. \ref{spectra_AS_v5}. 
In the fitting analysis for the spectrum at 1.9 K of the $x$ = 0.14 sample, we added a rotation component, $e^{-{{\it\lambda}_{3}}t}\cos(2{\pi}ft+\phi)$, to Eq. (\ref{eq:Eq_1}), where $\lambda_3$, $f$, and $\phi$ are damping rate, frequency, and phase of the muon spin precession.

	\begin{figure}[tb]
	\begin{center}
	\includegraphics[width=78mm]{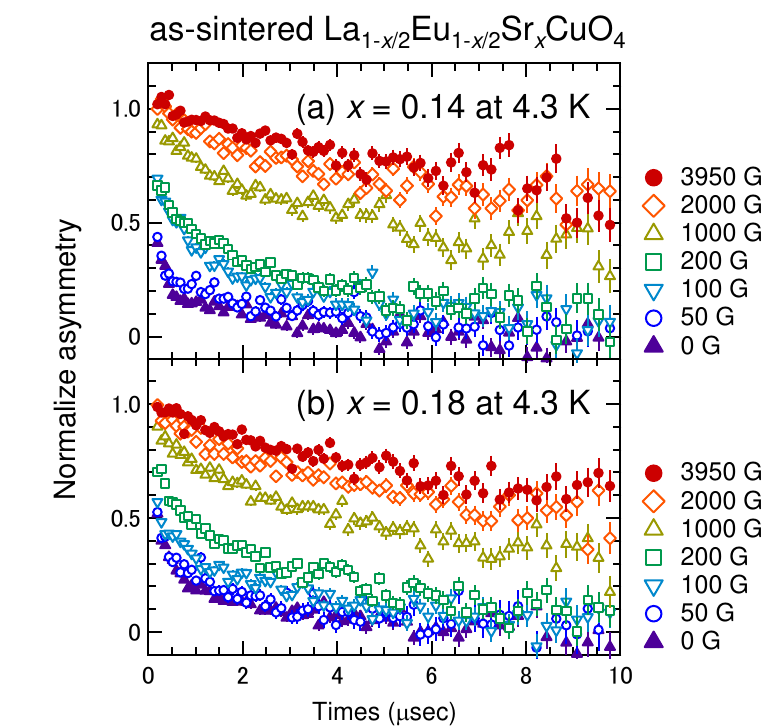}
	\caption{(Color online) Longitudinal-field $\mu$SR time spectra in as-sintered La$_{1-x/2}$Eu$_{1-x/2}$Sr$_x$CuO$_4$ with (a) $x=0.14$ at 4.3 K and with (b) with $x = $ 0.18 at 4.3 K measured in RAL.}
	\label{spectra_AS_LF_v3}
	\end{center}
	\end{figure}

Figure \ref{spectra_AS_Prm_v6}(a) presents the evaluated $\lambda_1$ as a function of temperature. Upon cooling, $\lambda_1$ grows at low temperature in all AS samples. 
Similarly, $A_{\rm 2}$ and $\lambda_2$ for all the measured samples increase with decreasing the temperature. (See Figs. \ref{spectra_AS_Prm_v6}(b) and \ref{spectra_AS_Prm_v6}(c).) This suggests that the (quasi) static internal magnetic field appears at muon sites due to the development of Cu spin correlation. 
%
%
To evaluate the onset temperature for the development of Cu spin correlation, we defined $T_{\rm m}$ as the intermediate point between the lowest temperature where  $A_{\rm 2}$ = 0 ($T_{\rm 1}$) and the first temperature where $A_{\rm 2}$ has a finite value upon cooling ($T_{\rm 2}$). (See Fig. \ref{spectra_AS_Prm_v6}(b).) Although the evaluation of the real onset temperature from a limited number of data is difficult, the $T_{\rm m}$ for a wide $x$ range shows a weak $x$-dependence within the accuracy. In Fig. \ref{phase_diagram_AS_v2}(a), $T_{\rm m}$ is plotted together with $T_{\rm sg}$, exhibiting the magnetic phase diagram for AS LESCO. 
Slightly higher $T_{\rm m}$ than $T_{\rm sg}$ likely originates from the different time windows of the probes to measure the slowing down phenomenon. This is consistent with the appearance of the SG phase~\cite{Wakimoto00, Enoki11, Fujita2008b,He2011,Suzuki2016}.


\subsection{Longitudinal-field $\mu$SR measurements}

The magnetic state of AS LESCO at low temperatures was further investigated by LF-$\mu$SR measurements to clarify the existence/absence of static magnetism. Figure \ref{spectra_AS_LF_v3}(a) shows the field dependence of LF-$\mu$SR time spectra for the $x=0.14$ sample at 4.3 K. Below 1000 Gauss, the depolarization of the spectra at longer time ranges was upwardly shifted with applying the magnetic field, whereas the field effect on the spectra in the vicinity of $t$ = 0 showing the rapid depolarization was negligible. With further increasing the field above 2000 Gauss, the overall spectra from $t$ = 0 were shifted; therefore, a static magnetic field lower than $\sim$2000 Gauss was present at the muon stopping site. The successive shift of the overall spectrum with the application of fields above 2000 Gauss suggests the existence of a fast fluctuating local field as well. 
Therefore, this sequential field dependence indicates the coexistence of static magnetic ordered and dynamically fluctuating spin states in the $x$ = 0.14 sample at 4.3 K. 
Similar field dependence of the LF-$\mu$SR time spectra was observed for the $x=0.18$ sample at 4.95 K, as seen in Fig. \ref{spectra_AS_LF_v3}(b), suggesting that the appearance of disordered magnetic state is common for the AS LESCO with a wide $x$ range. 

\section{Discussion}

\subsection{Magnetic and superconducting phase diagram}

\begin{figure}[tb]
	\begin{center}
	\includegraphics[width=82mm]{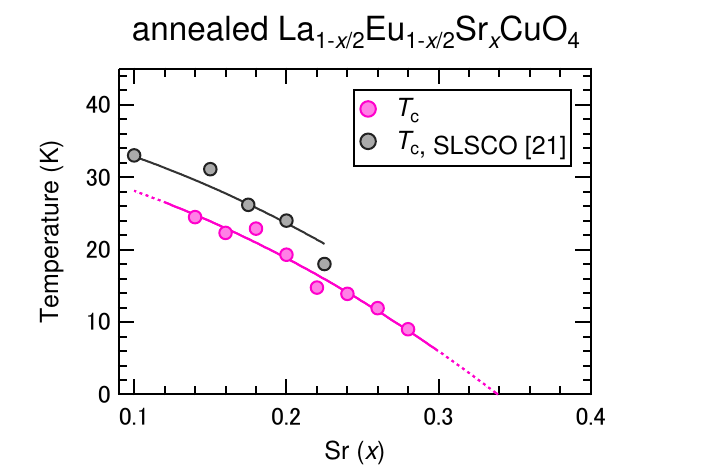}
	\caption{(Color online) Sr concentration dependences of $T_c$ in the oxidation annealed La$_{1-x/2}$Eu$_{1-x/2}$Sr$_x$CuO$_4$ and  one. Solid lines are guides to the eyes. $T_c$ for SmLa$_{1-x}$Sr$_{x}$CuO$_{4}$ (SLSCO) was taken from Ref.\cite{Kakeshita2009} is also plotted for reference.}
	\label{phase_diagram_AN}
	\end{center}
	\end{figure}

\begin{figure}[tb]
	\begin{center}
	\includegraphics[width=82mm]{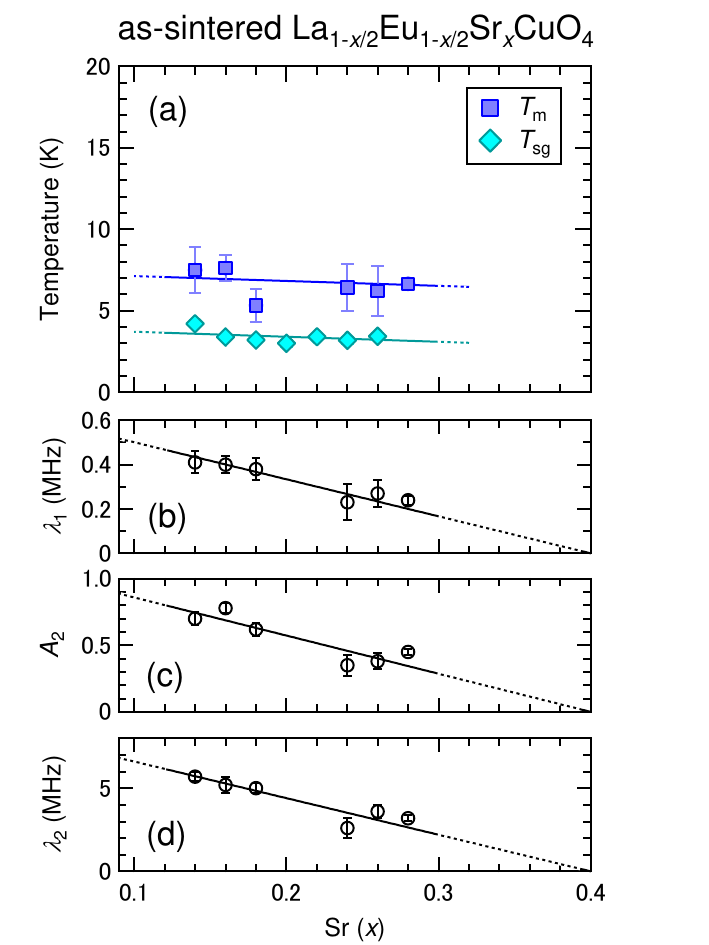}
	\caption{(Color online) (a) Sr concentration dependences of $T_{\rm m}$ and $T_{\rm sg}$ for as-sintered La$_{1-x/2}$Eu$_{1-x/2}$Sr$_x$CuO$_4$. Solid lines are guides to the eyes. Sr concentration dependence of (b) $\lambda_1$, (c) $A_2$, and (d) $\lambda_2$ at 4.5 K.}
	\label{phase_diagram_AS_v2}
	\end{center}
	\end{figure}

Our present study revealed that the oxidation annealing dramatically changes the electronic state, that is, the ground state of LESCO varies from SG to SC state through annealing for the first time. 
Figure \ref{phase_diagram_AN} and \ref{phase_diagram_AS_v2} (a) summarize $T_{\rm c}$ for the OA LESCO, and $T_{\rm m}$ and $T_{\rm sg}$ for AS LESCO. $T_{\rm c}$ for T*-type SmLa$_{1-x}$Sr$_{x}$CuO$_{4}$ (SLSCO) is also plotted as a reference~\cite{Kakeshita2009}. 
In the present study, the $x$ = 0.14 sample showed the maximum $T_{\rm c}$ and $T_{\rm c}$ decreased monotonically with increasing $x$, indicating that the present samples are located in the OD region. An extrapolation of the $x$-dependence of $T_{\rm c}$ yields $T_{\rm c}$ = 0 at $x$ $\sim$ 0.34 ($x_{\rm c}$), which is comparable to the critical value for the disappearance of the SC phase in T-type LSCO~\cite{Tanabe2005}. 
Absence of magnetic order in the present SC LESCO is consistent with the results for La$_{2-x}$Sr$_x$CuO$_4$ ($x$ = 0.15--0.20)\cite{Adachi2008}, Bi$_2$Sr$_{2-x}$La$_x$CuO$_{6+\rm{\delta}}$ (hole concentration, $p {\sim}$ 0.15--0.20) \cite{Russo2007}, and Bi$_{1.76}$Pb$_{0.35}$Sr$_{1.89}$CuO$_{6+\rm{\delta}}$ ($p>$ 0.09) \cite{Miyazaki2010}, and therefore, is a common feature of optimally-doped and overdoped (OD) single layer cuprate superconductors. 

On the contrary, the SG region in LESCO is quite broad against Sr concentration compared to that (0.02 $\leq$ $x$ $\leq$ 0.05) in LSCO and Bi$_{2+x}$Sr$_{2-x}$CuO$_{6+\delta}$, although $T_{\rm sg}$ in these single-layer cuprates is comparable~\cite{Wakimoto00, Enoki11, Enoki13}. 
Almost a constant $T_{\rm m}$ and $T_{\rm sg}$ against $x$ in the present AS LESCO implies the existence of magnetism far beyond $x_{\rm c}$. Consistently, the critical $x$ value for $A_{\rm 2}$ = 0 and $\lambda_{\rm 2}$ = 0 is estimated to be $\sim$0.4 from the extrapolation of $x$-dependence of $A_{\rm 2}$ and $\lambda_{\rm 2}$ at 4.5 K. (See Fig. \ref{phase_diagram_AS_v2}(c)--\ref{phase_diagram_AS_v2}(d).) Note that for each sample we evaluated $\lambda_{\rm 1}$, $A_{\rm 2}$ and $\lambda_{\rm 2}$ at 4.5 K from the interpolation or extrapolation of temperature dependence of values shown in Fig. \ref{spectra_AS_Prm_v6}(b)--\ref{spectra_AS_Prm_v6}(d).



\subsection{Origin of variation in magnetism due to oxidation annealing}

Here, we discuss the origin of the variation in magnetism due to oxidation annealing in T*-type LESCO. 
First, the semiconducting nature of AS samples indicates that the magnetism at low temperature corresponds to the localized Cu spins. 
Furthermore, as mentioned in the Introduction, the oxygen vacancies existing in the AS samples could be repaired through annealing~\cite{Sawa1989, Izumi1989, Border1990}. 
Considering the structural change of oxygen ions within the localized spin picture, following are the possible origins of disappearance of magnetic order due to oxidation annealing; 
 
 1) Increase of hole concentration,
 
 2) Removal of chemical disorder.

In case 1, 
oxidation of the sample causes the increase in the hole concentration due to the charge neutrality, and the carrier doping weakens the static spin correlations. 
The variation of carrier density due to annealing is indeed reported for T'-type cuprate~\cite{Song17, Horio18, Asano18}. The increased number of oxygens by the annealing ($y$) is $\sim$0.024 per formula unit for the present samples, suggesting that the additionally doped holes of 2$y$ at a fixed Sr concentration is 5\% at most. However, even though the Sr concentration, which corresponds to hole concentration, increases by 5\% in the AS LESCO, the system remains magnetic state and does not show superconductivity. Therefore, hole-doping by annealing is not the direct origin of the disappearance of magnetism. 

Regarding case 2, the chemical disorder potentially slows down the spin fluctuations and enhances  static magnetism. 
Considering the experimental fact that the magnetic properties concomitantly develop with the rapid increase of resistivity below a comparable temperature ($\sim$5--10 K), the suppression of carrier mobility is related with the stability of the magnetism. Such carrier localization at low temperature can be realized due to random potential around the disorders. Therefore, the removal of chemical disorders could result into the disappearance of the magnetism through the delocalization of carriers. 

There are three possible oxygen sites, which induces random potential on the CuO$_2$ plane~\cite{Sawa1989, Izumi1989, Border1990}: (i) apical oxygen site, (ii) oxygen site in the CuO$_2$ plane and (iii) interstitial oxygen site. 
The elongation of c-lattice constant after oxidation annealing suggests the insertion of oxygen between the CuO$_2$ planes, supporting the case (i) and (iii). 
It was indicated that the effect of local disorder on $T_{\rm c}$ was more marked in T*-type cuprates than in other single-layer ones, because of the shorter distance between Cu and apical oxygens~\cite{Kakeshita2009}. 
By the same reason, the disorder due to apical oxygen in T*-type cuprates could more markedly affect the spin and electric correlations on the CuO$_2$ planes. 
The effect of disorder at the interstitial oxygen sites, which are away from the CuO$_2$ plane, could be negligible. Thus, the main reason for the variation in magnetism due to oxidation annealing is most likely related to the apical oxygen. 
From a viewpoint of carrier localization, the number of effective mobile carriers as well as the mobility are considered to be lower than those in LSCO with the comparable $x$. The larger localization effect of hole carriers possibly result into the broad SG phase in AS LESCO.

\section{Summary}

We performed the first systematic $\mu$SR study of magnetism in T*-type cuprates using La$_{1-x/2}$Eu$_{1-x/2}$Sr$_x$CuO$_4$ with 0.14 $\leq$ $x$ $\leq$ 0.28. It was found that all AS semiconducting samples exhibit SG-like magnetism below $\sim$4 K. The internal magnetic field and the magnetic volume fraction concomitantly grow with the increase of electrical resistivity upon cooling, indicating a close relation between the magnetism and the mobility of carriers. The magnetic phase is robust against $x$ but fully suppressed in the  OA  samples, suggesting the competitive relation between disordered magnetic and SC states. This significant variation in the ground state of T*-type La$_{1-x/2}$Eu$_{1-x/2}$Sr$_x$CuO$_4$ by oxidation annealing yields a unique opportunity for the study of the interplay between spin correlations and superconductivity. 

\section*{Acknowledgments}

The $\mu$SR experiments at the RIKEN-RAL Muon Facility in the Rutherford Appleton Laboratory (Proposal No.RB1670580) and at the Materials and Life Science Experimental Facility of J-PARC (Proposal Nos. 2015MP001 and 2018B0324) were performed under user programs. 
We greatly thank the RIKEN-RAL and the J-PARC staff for their technical support during the experiments, and Y. Ishikawa for his contribution in administrative work for IMSS.
We also acknowledge helpful discussions with T. Adachi, K. M. Kojima, and Y. Miyazaki.
This work was supported by MEXT KAKENHI, Grant Numbers 16H02125,  the IMSS Multiprobe Research Grant Program, and IMSS Quantum Beam Research Grant.
		

\end{document}